\documentstyle[12pt]{article}
\date{}
\newcommand{\be}{\begin{equation}}
\newcommand{\ee}{\end{equation}}
\newcommand{\ba}{\begin{eqnarray}}
\newcommand{\ea}{\end{eqnarray}}

\begin{document}
\begin{flushright}
Preprint PITHA-94-23\\
July 1994 \qquad
\end{flushright}

\vspace{1.5cm}

\begin{center}
{\Large\bf Fine tuning in One-Higgs and Two-Higgs\\
 Standard Models}\\
\vspace{10mm}

{\large A. A. Andrianov
\footnote{On leave of absence from the Laboratory for Particle and
Nuclear Theory, Institute of Physics,
University of Sanct-Petersburg,
198904 Sanct-Petersburg,
Russia,\quad E-mail: andrianov1@phim.niif.spb.su},
\quad R. Rodenberg}\\

\medskip

{\small\it III. Physikalishes Institut\\
Abteilung f\"ur Theoretische Elementarteilchen Physik\\
RWTH Aachen, Physikzentrum, D-52056 Aachen, BRD}\\

\medskip

{\large and}\\

\medskip

{\large N. V. Romanenko}\\

\medskip

{\small\it Department of  Neutron Research,\\
Petersburg Nuclear  Physics Institute,\\
 188350, Gatchina, Russia}
\end{center}

\vspace{1cm}

\noindent
{\bf Abstract}\\

\medskip

 The fine-tuning principles are examined to
predict the top-quark and Higgs-boson masses. The modification of the
Veltman condition based on the
compensation of vacuum energies
 is developed. It is implemented in the Standard Model and in its minimal
 extension with two Higgs doublets.
The top-quark and Higgs-boson couplings are fitted in the SM for
the lowest ultraviolet scale where the fine-tuning can be stable under
rescaling. It yields
the low-energy values $m_t \simeq 175 GeV;\quad m_H \simeq 210 GeV$.

\newpage
\noindent
{\bf 1. \quad Introduction}\\

\medskip

The Standard Model (SM) describes
the strong and electroweak
particle interactions for a whole range of
energies which have been
available in experiments [1]. Still, there is a number of
well-known problems that are
to be
resolved in order to justify all the principles which the Standard
Model is based on. In particular,
the detection of top-quark is expected [2]  and
the discovery of scalar Higgs particle is wanted
[1,3].

In this connection, much effort has been taken to estimate their masses
for the purpose of understanding the
possible extensions of SM. Among others the minimal and natural
generalization
of the Standard Model which contains three generations and the basic
mechanism of mass generation is given by the extension of the Higgs
sector of the SM [4, 5]. For instance, the Minimal Supersymmetric SM
entails two Higgs doublets at low energies [6]. Thus the search for
relations between the many Higgs-field dynamics and the masses
of $t$-quark and Higgs-bosons may give the selection rule for a particular
model beyond the SM as well as for its acceptable parameters [3, 5].

There exist a few phenomenological principles
within the SM and its minimal extensions which make it possible
to determine relations between top-quark and
Higgs-boson masses. These principles are based on the assumption
that the SM is actually an effective theory applicable for
low energies. Consequently, its coupling constants
and dimensional parameters absorb all the influence of high-energy degrees
of freedom and of very heavy particles as well. Of course, the form of
effective
action [7] generally depends on the
preparation procedure, but the low-dimensional part of
the ef\/fective action coincides with
the SM action and is  minimally sensitive to very
high energies. The decoupling of the high-energy dynamics responsible
for the parameter formation is manifested in the relations between
dimensional parameters and certain coupling constants.
The latter statement allows to formulate the following
phenomenological principles which could be realized
 in the quantum SM or in its minimal extensions.
\begin{itemize}
\item The strong fine tuning for the Higgs field parameters
(v.e.v and its mass) that consists in the
cancellation of large radiative contributions quadratic in ultraviolet scales
which bound the particle spectra in the effective theory (Veltman condition
[8-12]).
\item The strong fine tuning for vacuum energies [13]
that envisages the cancellation of
large divergencies quartic in ultraviolet scales which might effect
drastically the formation of the cosmological constant.
\item The RG stability of the cancellation mechanism under
change of ultraviolet scale of effective theory [10, 13].
\end{itemize}
Hereby we suppose that the consistent
preparation of an ef\/fective model is to provide the decoupling of
low-energy world from very high energies, and,
therefore for an appropriate choice
of ultraviolet scales neither large vacuum energies nor large
Higgs-boson masses
should not arise [13].
The ultraviolet scales of the effective action are supposed to simulate
the average masses of very heavy degrees of freedom which however should
not shift the vacuum energy apart from zero. The RG stability of
fine-tuning relations is a further consequence of decoupling and provides
their applicability within the entire energy range of effective SM. It
can be interpreted as the RG invariant reduction condition in space of
coupling constants in the spirit of [14].

In our paper we examine the compatibility of these principles
for the SM and for the two-Higgs model.
The vacuum-energy fine tuning when combined with others leads to
 predictions for the top-quark and Higgs-boson masses within the range
of validity of the Standard Model which are in the fair correspondence
to the recent experiments [2].\\

\bigskip

\noindent
{\bf 2.\quad Veltman's fine-tuning}\\

\bigskip

Let us remind that the scalar sector in the Weinberg-Salam theory
contains the quadratic divergences in the tadpole diagrams and in the scalar
particle self-energy.  In order to fit the finite masses
the rule of cancellation for
 quadratic divergences, the fine-tuning,
[8] was proposed. This cancellation
holds only if the fermion and boson loops are tuned due to specific
values of coupling constants. At the one
loop level the condition:
\be
(2M^2_W+M^2_Z+m^2_H)=\frac{4}{3}
\sum_{flavors, \; colors} m^2_f     \label{eq:Velt}
\ee
removes quadratic divergences both from the Higgs-field v.e.v. and the
Higgs boson self-energy if \underline{the universal momentum cutoff}
 is implemented for all the fields.
{}From (\ref{eq:Velt}) it is easy to see that if the $t$-quark is the only
heavy fermion, $m_t \geq 70 GeV$ bound should be fulfilled.

The original Veltman condition is however not stable under rescaling
since its renorm-derivative cannot vanish simultaneously for any
choice of the cutoff $\Lambda$ below the Plank scale [10],
i. e. the related cancellation of quadratic divergences can be
provided only at a selected scale.

On the other hand, the usage of the universal scale for all bosons
and fermions roughly implies the existence of large symmetry involving
all the observable particles in one
multiplet and therefore is not well motivated
within the framework of effective theory.\\

\bigskip

\noindent
{\bf 3.\quad Vacuum fine-tuning in the Standard Model}\\

\bigskip

Let us consider the SM as a low-energy limit of a more
fundamental theory and suppose that only one heavy fermion, t-quark
takes part in its dynamics within the selected energy range. Respectively,
 we
neglect the masses of all lighter fermions. When there is no expected
supersymmetry (below the scale $\Lambda_{new}$
where the SM is valid) we apply different
scales for the design of SM-effective action for bosons $\Lambda_{B}<
\Lambda_{new}$ and for fermions $\Lambda_{F}< \Lambda_{new}$. Among
bosons the universal scale is introduced in order not to induce
artificially the
explicit breaking of a Grand
Unification symmetry. In its turn,
the universal scale for fermions confirms the horizontal symmetry in the
ultraviolet region.

We require for the SM that very large contributions (leading
divergencies) into dimensional physical parameters
should be suppressed (which is equivalent to the absence
of their strong scale dependence) and,
in addition to the strong fine-tuning,
the cancellation of contributions into vacuum energy should take place,
\ba
T_{\mu \nu}&\sim& g_{\mu \nu}E_{vac}\approx 0;\quad
(4N_F + 2N_{\nu})\,\Lambda_F^4 = (2N_B + N_S)\,\Lambda_B^4;\nonumber\\
\alpha^2 &=& \frac{\Lambda_b^4}{\Lambda_F^4}=\frac{4N_F + 2N_{\nu}}{2N_B+N_S}
= \frac{45}{14};
\ea
where $N_F = 21$ is the number of flavor and color degrees of
freedom for three generations of massive fermions, $N_{\nu}$ is the
number of (Weyl or Majorana) neutrinos,
 $N_B = 12,\, N_S= 4$ are numbers of
flavor and color degrees of freedom for vector and scalar bosons,
respectively.  If neutrinos were massive Dirac particles then one should
replace $N_{\nu} \rightarrow 2N_{\nu}$ which yields $\alpha^2 = 24/7$.
Since the problem of neutrino mass is not yet solved we shall keep in
mind both possibilities.

Respectively the strong fine-tuning condition at the one-loop
level reads:
\be
4m^2_t=\alpha(2M^2_W+M^2_Z+m^2_H);\quad \alpha \simeq 1.793
\quad\mbox{or}\quad 1.852. \label{eq:xx}
\ee
 Taking into account the ef\/fects of all loops one obtains the integral
condition of Veltman type:
\be
\int_{v}^{\Lambda_{F}}  \frac{d^4k}{k^2} 4g_t^2=
\int_{v}^{\Lambda_{B}}\frac{d^4k}{k^2}\biggl\lbrace \frac32 g^2 +
+ \frac12 g'^2+
2\lambda \biggr\rbrace + O\left(v^2 \ln(\Lambda^2/v^2)\right);
\ee
where the conventional denotations for the electroweak $g,\,g'$,
Higgs-quartic,
$\lambda$ and the Yukawa $t$-quark, $g_t$ coupling constants are used.
When integrating by  parts one can conclude that the leading contribution is
the modified Veltman condition at the scale $\Lambda$. The latter one is
supplemented in the two-loop approximation
with its renorm-derivative (having small combinatorial factor) and so on.
At the one-loop level of accuracy, all the renorm-derivatives
except the first one are zero.

 Let us prolongate the validity of the fine tuning to the entire
 energy range below
 the effective scale $\Lambda$ and impose the requirement
 of its weak dependence of scale ,
 \be
f \equiv 4 g_t^2 - 2 \alpha (\lambda +A)=0;\quad
Df \equiv 16 \pi^2 \frac{\partial f}{\partial \tau} =0;\quad
\tau = \ln\frac{\Lambda}{v_0}. \label{MV1}
\ee
The explicit form of the stability condition is:
\begin{eqnarray}
Df  = && 8g_t^2 \left[ \frac{9}{2}g_t^2-8g_3^2-\frac{9}{4}g^2-
\frac{17}{12}g'^2 \right]
 - 24 \alpha \left[ \lambda^2 + (g_t^2-A)\lambda +B - g_t^4 \right]
   \nonumber \\
  && - \frac{\alpha}{2} (-19g^4 + \frac{41}{3}g'^4), \label{MV2}
 \end{eqnarray}
in Eqs.(\ref{MV1}), (\ref{MV2}) the following denotations are used:
\be
A \equiv \frac{3}{4}g^2+\frac{1}{4}g'^2;\quad
B=\frac{1}{16}g'^4+\frac{1}{8}g^2g'^2 +\frac{3}{16}g^2.
\ee
After elimination of $\lambda$
the equation for the $t$-quark Yukawa coupling constant reads,
\be
k_1 g_t^4 + k_2 g_t^2 + k_3 =0 , \label{finsm}
\ee
where the coef\/f\/icients are:
$$k_1=12 \left(-2\alpha + 1 + \frac{8}{\alpha}\right)
\approx 22.5 \quad(19.4);$$
\be
k_2= 64 g_3^2 + \frac{16}{3} g'^2 -24 (\alpha  + 5) A;\quad
k_3 =\frac{\alpha}{9}\left(352 A^2 + 76 A g'^2 + 61 g'^4\right)
\ee
Numerically the existence of a solution is very sensitive  both to the
value of $\alpha\simeq 1.80 (1.85)$
and to the value of the strong coupling constant
$\alpha_3 = g^2_3/4\pi$. The averaged value
of $\alpha_3$ is taken from [15] as
$ \alpha_3(M_Z) = 0.118 \pm 0.007$.

It can be found from eq.(\ref{finsm}) that the solution  for $g_t^2$
exists only for the effective scale $\Lambda \sim 10^{15} GeV$
when EW coupling
constants approach to their GUT values $g^2_3 \sim g^2 \sim (5/3)g'^2$.
The low-energy value for the
Yukawa coupling constant $g_t$ is delivered by the renormalization-group
flow,
$$Dg_t =  g_t \left ( \frac29 g^2_t - \frac{17}{12} g'^2 -\frac94 g^2
-8 g^2_3\right).$$

The low energy values of $m_H$ are obtained with the help of the IR
quasi-fixed
point in the RG-equation for the Higgs self-coupling [16]
$$
D \lambda=
12 \left( \lambda^2 +(g_t^2-A) \lambda -g_t^4 +B \right).$$
It happens [16] that $\lambda(\tau)$ tends to
the Hill's quasi-fixed point  ($D\lambda
=0$) in the wide intermediate energy range for
\underline{any} boundary conditions at high energies.

The stability condition (6) ensures the strong f\/ine-tuning both
to two-loop level and numerically and leads at the EW scale to the
predictions,
\ba
\left\{
\begin{array}{l}
m_t (v) = 177 \pm 5\, GeV,\\
m_H (v) = 213 \pm 10\, GeV.
\end{array} \right.
\ea
The dependence of the neutrino degrees of freedom is rather weak
and is included into the error bar.

The total theoretical error can be estimated from
the evaluation of the 2-loop contributions [10] into
$\beta$-functions and from the error bar in the determination
of $\alpha_s$.
One can check up that for these mass values the modified
Veltman condition holds with good accuracy
both for the effective bound of the Standard Model $\Lambda \approx 5\cdot
10^{15} GeV$ and for the vector boson mass scale $\mu \approx 100 GeV$.
This means that the modified Veltman equation does not
depend on the rescaling for the wide range of energies.
The predictions are within the accepted range
of different experimental and theoretical
bounds [1, 2].

Nevertheless one could find the application of the SM particle
content and of its renormalization group up to such huge
energies not well motivated and therefore let us examine in what
cases the fine-tuning can implemented  in the minimally extended
SM with two-Higgs doublets.\\

\bigskip

\noindent
{\bf  4.\quad Vacuum fine-tuning in the Two-Higgs Model}\\

\bigskip

We consider the Two-Higgs model with the potential,
 $$V(H_1, H_2)= \frac{\lambda_1}{2}(H_1^{\dagger}H_1)^2 +
 \frac{\lambda_2}{2}(H_2^{\dagger}H_2)^2
 + \lambda_3(H_1^{\dagger} H_1)(H_2^{\dagger}H_2)+ $$
\be
 +\lambda_4(H_1^{\dagger}H_2)(H_2^{\dagger}H_1)
 +\frac{\lambda_5}{2}\left( (H_1^{\dagger}H_2)^2 +(H_2^{\dagger}H_1)^2
 \right),
\ee
with two Higgs doublets,

\be
H_1= \frac{1}{\sqrt{2}}\left(
 \begin{array}{c}
 0 \nonumber \\ v_1
  \end{array}
   \right)
   + \frac{1}{\sqrt{2}} \left(
 \begin{array}{c}
 \phi_2 +i \phi_1 \nonumber \\ \chi_1- i \phi_3
  \end{array}
   \right)    ; \quad
 H_2= \frac{1}{\sqrt{2}}\left(
 \begin{array}{c}
 0 \nonumber \\ v_2
  \end{array}
   \right)
   + \frac{1}{\sqrt{2}} \left(
 \begin{array}{c}
h_2 +i h_1 \nonumber \\ \chi_2- i h_3
  \end{array}
   \right),
\ee
herein $v_{1,2}$ are v.e.v. of Higgs fields.
This potential possesses the discrete symmetry,
 $H_1 \rightarrow H_1$; $H_2 \rightarrow -H_2$.

 The constants in the Higgs potential should make it bounded from below
 that is guaranteed when
 \be
 \lambda_{1,2}> 0, \quad\sqrt{\lambda_1 \lambda_2} > -\lambda_3 -
 \lambda_4 + |\lambda_5|. \label{vst}
 \ee
  We require also that the vacuum configuration conserve electric charge,
 $\lambda_4<0$. The choice of the $H_2$ phase may be always done so that
 $\lambda_5<0$.

 Let us neglect small Yukawa coupling constants for two first
 quark generations as well as for all leptons. Then the remaining
 interaction of Higgs fields with $b$- and $t$-quarks reads:
\be
L_{Yuk}= -\overline{Q}_L(G_1^b H_1 +G_2^b H_2)  b_R
 -\overline{Q}_L i \tau^2(G_1^t H_1^* + G_2^t H_2^*) t_R \;+ \;(h.c.).
\ee
The realistic ratio $m_t/m_b << 1$ can be produced in several ways.\\
1) The mass hierarchy can be treated as the consequence
of a hierarchy of coupling constants,
$$G_1^d=G_2^u=0, \quad G_2^d \equiv g_b << G_1^u \equiv g_t,$$ while both
v.e.v.'s are comparable in their magnitudes.
It will be shown that in this case the cancellation of the
quadratic divergences cannot be supplied with the one-loop RG invariance
at every scale.\\
2)The mass difference may be caused by the hierarchy of v.e.v's:
$m_b/m_t= v_2/ v_1 << 1$,
$$G_1^u= G_2^d \equiv F;\quad  G_2^u=G_1^d=0.$$
For such a choice the global
right symmetry arises in the limit $g' \rightarrow 0$.
Then it happens to be possible  to
cancel both vacuum energies  and quadratic divergencies and
furthermore to implement the one-loop RG invariance of these conditions.\\
3) The small value of  $m_b$ can be delivered by a tuning of Yukawa constants
$F$ and $G$,
$$G_1^u= G_2^d \equiv F; \quad G_2^u= G_1^d\equiv G \not=0.$$
(again with the global right symmetry when hypercharges  are neglected)
This or  a more general model seems to be less natural since
it is difficult to avoid non-conserving strangeness
neutral currents when the Cabibbo mixing is taken into account.
By this reason
we restrict ourselves with the analysis of two first cases.\\

Let us derive the vacuum fine-tuning conditions for the Two-Higgs model.
The vacuum-energy cancellation  reads:
$$\alpha^2 \equiv \frac{\Lambda_b^4}{\Lambda_F^4}=
\frac{4N_F+ 2N_{\nu}}{2N_B+N_S} = \frac{45}{16}\quad (\mbox{or}\quad 3);\quad
\alpha \approx 1.677 \quad(\mbox{or}\,\,\, 1.732).$$

There are two conditions of Veltman type
 corresponding to two mass terms in the lagrangian:
$$-m_1^2 (H_1^+H_1)) -m_2^2 (H_2^+H_2).$$
 In the cases 1) and 2) the modified Veltman conditions
 take the following form:
\be \left\{ \begin{array}{l}
f_t \equiv 4g_t^2- \alpha [ \frac{3}{2}g^2 +\frac{1}{2}g'^2 + 2 \lambda_1
+\frac{4}{3} \lambda_3 + \frac{2}{3} \lambda_4]    \nonumber \\
f_b \equiv 4g_b^2- \alpha [ \frac{3}{2}g^2 +\frac{1}{2}g'^2 + 2 \lambda_2
+\frac{4}{3} \lambda_3 + \frac{2}{3} \lambda_4]
\end{array} \right. \label{2hft}
\ee
Their one-loop RG invariance is provided if
\be
Df_t=0; \quad Df_b=0, \label{2hrg}
\ee
where the definitions from Sec.3 are employed.

  Let us investigate RG-invariance of the modified Veltman conditions.
We have four independent equations (\ref{2hft}), (\ref{2hrg}).
Taking the difference between the first and the second equation
in (\ref{2hft})
 and in (\ref{2hrg}) one obtains:
$$4(g_t^2- g_b^2)- 2\alpha(\lambda_1-\lambda_2)=0;$$
$$36(g_t^4- g_b^4) - (g_t^2 - g_b^2)(64g_3^2+18g^2+\frac{22}{3}g'^2)-
4g'^2 (g_t^2+g_b^2)-$$
\be
-24\alpha[(\lambda_1-\lambda_2)(\lambda_1+\lambda_2-
\frac{3}{4}g^2+ \frac{1}{4}g'^2)+
g_t^2(\lambda_1-g_t^2) - g_b^2(\lambda_2-g_b^2)]=0. \label{dif}
\ee

In the case 1) the terms containing $g_b$ can be neglected.
 Then subsituting the first equation
 into the second and excluding $\lambda_1$
 one has:
 \be
 \lambda_2=\frac{1}{12(\alpha+4)}\left\{-(6 +
 \frac{48}{\alpha}- 12 \alpha) g_t^2-
 32g_3^2+ 9g^2+\frac{1}{3}g'^2\right\}.
           \ee   \label{eq:1}
 It is easy to see that $\lambda_2 < 0$ for $\alpha < 2.3$,
 independently of the energy scale. But it should be positive in order
 to yield the potential bounded from below [4].
 Thus the RG-invariant fine-tuning cannot be implemented when
 $g_b << g_t$.

In the case 2)
the modified Veltman conditions coincide for
 both v.e.v.'s. A little difference appears for the renorm-derivatives
 due to the different hypercharges but it may be neglected for the
 numerical calculations: this will give the systematic error
 in the definition of $m_t$ about $\pm2\, GeV$.

 Let us adopt the global right symmetry of the Higgs potential and
 select out $\lambda_1 \simeq \lambda_2 \simeq \lambda_3
 \equiv \bar\lambda$. Again a little discrepance exists in the RG
 equations for $\lambda_{1,2}$ and $\lambda_3$ (see Appendix).
 In what follows
 we interpolate the RG equations for $g_t \simeq g_b \equiv F$ and
 $\lambda_{1,2,3}
 \equiv \bar\lambda$ by their averages,
 \ba
 DF &=& F \left( -(8g_3^2 + \frac{9}{4}g^2 + \frac{11}{12}g'^2)
 + 5 F^2 \right);\nonumber\\
D\bar\lambda &=& 12 \bar\lambda^2 + 4\lambda_3^2 + 4 \lambda_3 \lambda_4
+2 \lambda_4^2 +2 \lambda_5^2 \nonumber\\
&&-3\bar\lambda(3g^2+g'^2) +
12\bar\lambda F^2 +\frac{9}{4}g^4 + \frac{3}{4}g'^4
- 12 F^4.
\ea
In this case the fine-tuning conditions,
  \be \left\{ \begin{array}{l}
  f\equiv 4F^2- \alpha[\frac{3}{2}g^2+\frac{1}{2}g'^2+
  \frac{10}{3}\bar\lambda+\frac{2}{3}\lambda_4]=0; \nonumber \\
  Df=0;
  \end{array} \right. \ee
can be reduced to the equation for the Yukawa coupling constant
when excluding $\bar\lambda$ from the first equation:
\be
k_1F^4+k_2F^2 +k_3=0. \label{eqF}
\ee
Here:
\ba
k_1&=& 32 \alpha -8 - \frac{384}{5\alpha} \simeq -0.13 \quad (3.09) ;
\nonumber\\
k_2&=& -64g_3^2 + \left(\frac{378}{5} + 24\alpha\right)g^2 +
\left(\frac{358}{15} + 8\alpha \right)g'^2;\nonumber\\
k_3&=&-\alpha\left( \frac{114}{5}g^4 + \frac{91}{5}g^2 g'^2
+ \frac{61}{5}g'^4
+ \frac{36}{5}\lambda_4^2+ 12 \lambda_5^2\right) < 0.
\ea
It is evident from the signs of $k_{1,3}$ that eq.(\ref{eqF}) has one
positive solution for any constants $g_3, g, g', \lambda_{4,5}$ and
therefore for any values of the cutoff (in the contrast to the
Standard Model with one Higgs field). However  the value of $F$
becomes larger with increasing $\lambda_{4,5}$. On the other hand,
we are interested to find the set of parameters in the two-Higgs
potential which yields the minimal values of the $t$-quark mass.
Therefore let us provide the condition for Peccei-Quinn (PQ) symmetry
(which is RG-invariant):
 $\lambda_5=0$.
Besides one could minimize further on when imposing $\lambda_4 \approx 0$
at a particular scale (it is not RG invariant, see Appendix).
Below on the numerical estimations
 of minimal values for $m_t$ are presented for different energies.

\vspace{1cm}

\begin{center}
\begin{tabular}{||l|c|c|c|c||} \hline \hline
$\Lambda \, GeV$
& $``m_t(\Lambda,\nu_{Dir})"$ &$m_t(100 \, GeV, \nu_{Dir})$&
$ ``m_t(\Lambda,\nu_{Weyl})"$
&$m_t(100 \, GeV, \nu_{Weyl})$ \\ \hline
$10^{15}$        &  99 & 160 & 101 & 162 \\ \hline
$10^{14}$         &   101 & 161 & 104 & 163 \\ \hline
$10^{13}$         & 104   & 162 & 108 & 165 \\ \hline
$10^{12}$         & 108   & 164 & 112 & 167  \\ \hline
$10^{11}$         & 113   & 166 & 119 & 170 \\ \hline
$10^{10}$         & 120   & 170 & 127 & 174 \\ \hline
$10^{9}$          & 129   & 175 & 140 & 181 \\ \hline
$10^{8}$          & 143   & 182 & 163 & 194 \\ \hline
$10^{7}$          & 165   & 195 & 212 & 216  \\ \hline\hline
\end{tabular}

\vspace{5mm}

{\bf Table 1.}\quad Masses of the $t$-quark for $\lambda_{4,5} \simeq
0$ at a scale $\Lambda$.
\end{center}

\vspace{0.6cm}

In order to predict the real $m_t$ we use the RG flow
which is described approximately by the following equation,
$$F^2(\mu)= \left(\frac{g_3^2(\mu)}{g_3^2(\Lambda)}\right)^{8/7}
\cdot \frac{F^2(\Lambda)}{1+\frac{5F^2(\Lambda)}{g_3^2(\Lambda)}
\left[ \left(\frac{g_3^2(\mu)}{g_3^2(\Lambda)}   \right)^{1/7}-1 \right]}.$$
Predictions for the Higgs spectra can be found with help of
the quasi IR fixed points [17].
For the chosen scheme of couplings:
$m_+ \approx 200-205 \;GeV $; $m_1 \approx 225-230 \; GeV$;
$m_2 \approx 6-6.5 \;GeV  $; $m_p \approx 0\; GeV$ (PQ-symmetry).

If one imposes the modified fine-tuning relation at low energies, such
Higgs masses would lead  to $m_t\approx 173 \;GeV$. Thus one conclude
that the scale $\Lambda \simeq 10^9 GeV$ is preferable to implement the
RG-invariant fine-tuning with a good precision in the entire energy
range below this scale. It is surprising that the estimations for
the $t$-quark mass are close to the fine-tuning predictions
of the one-Higgs Standard Model and to the recent experimental
data [2].\\

\bigskip

\noindent
{\bf 5.\quad Conclusions.}\\

\bigskip

We have shown that in the Standard Model with one and two Higgs
doublets the selection rule
based on the vacuum fine-tuning can be implemented for the parameters
of $t$-quark and Higgs-boson
potentials. This is not possible in the original Veltman's formulation
since the coefficients $k_1$, eqs.(\ref{finsm}), (\ref{eqF})
in both models are negative and large. To our mind, the approximate
RG invariance is an important property of the fine-tuning conditions
which otherwise do not acquire the universal meaning. Therefore
we have developed the approach different from [18].
As well we cannot agree with the empirical application of
the original Veltman condition to the physics at the $W$-boson scale
[19] while
it is supposed to be valid at the scale of new physics. It could be
reasonable if this condition were RG-invariant.
In the two-Higgs model the vacuum fine-tuning can be realized when
the Yukawa coupling constants for $b$- and $t$-quarks are comparable
that means the hierarchy of v.e.v.'s for the Higgs fields.
Thereby the vacuum fine-tuning may give a resolution between the
different scenarios to generate the hierarchy of quark masses.

\bigskip

One of us (A.A.A.) is very grateful to the community of the III.
Physi\-ka\-li\-sches Institut of RWTH-Aachen and especially to
Prof. Rodenberg for the warm hospitality. We also thank the
German Research Foundation (DFG) for financial support of our
collaboration. A.A.A. is partially supported by the Russian
GRACENAS (Grant No. 2040).

\vspace{5mm}

\noindent
{\bf Appendix}\\

\bigskip

\noindent
The one-loop RG equations of the Two-Higgs model in Cases 1), 2) read:

$$D g'=7g'^3; \quad Dg=-3g^3; \quad Dg_3=-7g_3^3; \eqno{(A1)}$$

$$ Dg^t = g^t \left( -(8g_3^2 + \frac{9}{4}g^2 + \frac{17}{12}g'^2)
+\frac92 g_t^2 + \frac12 g_b^2 \right); \eqno{(A2)}$$

$$ Dg^b = g^b \left( -(8g_3^2 + \frac{9}{4}g^2 + \frac{5}{12}g'^2)
+\frac92 g_b^2 + \frac12 g_t^2 \right); \eqno{(A3)}$$

$$ D\lambda_1= 12 \lambda_1^2 + 4\lambda_3^2 + 4 \lambda_3 \lambda_4
+2 \lambda_4^2 +2 \lambda_5^2 $$
$$-3\lambda_1(3g^2+g'^2) +
12 \lambda_1 g_t^2 +\frac{9}{4}g^4 +\frac{3}{2}g^2g'^2+\frac{3}{4}g'^4
- 12 g_t^4; \eqno{(A4)}$$

$$ D\lambda_2= 12 \lambda_2^2 + 4\lambda_3^2 + 4 \lambda_3 \lambda_4
+2 \lambda_4^2 +2 \lambda_5^2 $$
$$-3\lambda_2(3g^2+g'^2) +
12 \lambda_2 g_b^2 +\frac{9}{4}g^4 +\frac{3}{2}g^2g'^2+\frac{3}{4}g'^4
- 12 g_b^4; \eqno{(A5)}$$

$$ D\lambda_3= (\lambda_1+\lambda_2)(6\lambda_3+2\lambda_4)+
4\lambda_3^2   +2 \lambda_4^2 +2 \lambda_5^2  $$
$$-3\lambda_3(3g^2+g'^2)+
6 \lambda_3 (g_t^2 + g_b^2) +\frac{9}{4}g^4 -\frac{3}{2}g^2g'^2
+\frac{3}{4}g'^4
- 12 g_t^2 g_b^2; \eqno{(A6)}$$

$$D\lambda_4= 2(\lambda_1 +\lambda_2)\lambda_4+
4(2\lambda_3+\lambda_4)\lambda_4 +8\lambda_5^2 $$
$$-3\lambda_4(3g^2+g'^2)
+6 \lambda_4 (g_t^2 + g_b^2) +3g^2g'^2  + 12 g_t^2 g_b^2; \eqno{(A7)}$$

$$D\lambda_5=\lambda_5 \left(2(\lambda_1+\lambda_2)
+8\lambda_3 +12 \lambda_4-3(3g^2+g'^2) +
6(g_t^2 + g_b^2)\right). \eqno{(A8)}$$

\vspace{0.5cm}

\noindent
{\bf References}\\

\medskip

\begin{itemize}
\item[1.] K. Hikasa et al., Particle Data Group: Phys. Rev.
D45 (1992)(Part 2)
\item[2.]  F. Abe et al.,CDF collab.:
FERMILAB Pub-94/097-E, CDF, 1994
\item[3.] R. Rodenberg: FERMILAB-PUB-93/-T, 1993;\,\,
FERMILAB-93/103-T, 1993
\item[4.] G. Kreyerhoff, R. Rodenberg:
Phys. Lett. B226 (1989) 323;\,\,
J. Freund, G. Kreyerhoff, R. Rodenberg:
Phys. Lett. B280 (1992) 267
\item[5.] N. V. Krasnikov, S. Pokorski: Phys. Lett. B288 (1992) 184.
\item[6.] N. Cabibbo, L. Maiani, G. Parisi,
R. Petronzio: Nucl. Phys.  B158 (1979) 295;\,\,
H. P. Nilles: Phys. Rep. 110 (1984) 1
\item[7.] K. Wilson and J.B. Kogut:  Phys. Rep. 12C (1974) 75
\item[8.] M. Veltman: Acta Phys. Polon. B12 (1981) 437
\item[9.]  Y. Nambu: EFI-90-46, 1990
\item[10.] I. Jack, D. R. T. Jones: Nucl. Phys. B342 (1990) 127;
\,Phys. Lett. B234 (1990) 321;\,\,
 M. B. Einhorn, D. R. T. Jones: Phys. Rev. D46 (1992) 5206;\,\,
 M. S. Al-sarhi, I. Jack, D. R. T. Jones Z. Phys. C55 (1992) 283.
\item[11.] M. Ruiz-Altaba, B. Gonzales, M. Vargas:
CERN-TH 5558/89, 1989;\,\,
M. Capdequi Peyranere, J. C. Montero, G. Moultaka:
Phys. Lett. B260 (1991) 138
\item[12.] P. Osland, T. T. Wu: Z. Phys. C55 (1992) 569, 585, 593;\,
Phys. Lett.  B291 (1992) 315
\item[13.] A. A. Andrianov, V. A. Andrianov, N. V. Romanenko:  in:
Proceedings of XIV Seminar on High Energy Physics Problems, Protvino,
Russia: Nauka, (1991) 284;\,\,
A. A. Andrianov, N. V. Romanenko:
Phys. Atom. Nucl. (Yad. Fiz.)  57 (1994) 509
\item[14.] J. Kubo, K. Sibold, W. Zimmermann: Nucl.Phys. B259 (1985)331;
\quad Phys.Lett. B220 (1989) 191
\item[15.] G. Altarelli: CERN-TH.6623/92, 1992
\item[16.] C. T. Hill: Phys. Rev. D24 (1981) 69;\,
FERMI-CONF 90/17 T., 1990
\item[17.] C. T. Hill, C. N. Leung, S. Rao:  Nucl.Phys. B262 (1985) 517
\item[18.] C. Newton, T. T. Wu: Z. Phys. C62 (1994) 253
\item[19.] G. Passarino: Z. Phys. C62 (1994) 229
\end{itemize}
\end{document}